\documentclass[12pt,fleqn]{article}
\usepackage{geometry}                
\usepackage{graphicx}
\usepackage{amssymb}
\usepackage{epstopdf}
\usepackage{hyperref}
\hypersetup{colorlinks,%
	citecolor=blue,
	}

\usepackage{latexsym}
\usepackage{marvosym}

\usepackage[mathcal]{euscript}
\usepackage{indentfirst}

\def\K3{\mathrm K3}

\def\normal#1{\,\!:\! #1 \!:\! \,}
\def\double #1{#1{\hbox{\kern-2pt $#1$}}}

\begin{document}

\vspace{-70pt}

\begin{flushright}
\makebox[0pt][b]{}
\end{flushright}

\vspace{10pt}

\center{{\LARGE The Covariant Superstring on K3}

\vspace{30pt}

{\large Osvaldo~Chand\'ia$,\hspace{-9pt}{}^{\small \mbox\Aries}$\
William~D.~Linch~{\sc iii}$,\hspace{-6pt}{}^{\small \mbox\Pisces}$\
and Brenno Carlini Vallilo$.\hspace{-4pt}{}^{\small \mbox \Libra}$ }
}


\center{
${}^{\small \mbox\Aries}${\em
Departamento de Ciencias, Facultad de Artes Liberales\\
\& Facultad de Ingenieria y Ciencias, \\
Universidad Adolfo Iba\~nez,\\
Santiago de Chile.}\\

\vskip 0.2in
${}^{\small \mbox\Libra,\mbox\Pisces}${\em
Departamento de Ciencias F\'isicas,\\
Facultad de Ciencias Exactas,\\
Universidad Andres Bello,\\
Santiago de Chile.}
}
\vspace{10pt}

\abstract{
We compactify the pure spinor formalism on a K3 surface. The pure spinor splits into a six-dimensional pure spinor, a projective superspace harmonic, and 6 non-covariant variables. A homological algebra argument reduces the calculation of the cohomology of the Berkovits differential to a ``small'' Hilbert space which is the string-theoretic analogue of projective superspace. The description of the physical state conditions is facilitated by lifting to the full harmonic superspace, which is accomplished by the introduction of the missing harmonics as non-minimal variables. Finally, contact with the hybrid formalism is made by returning to the small Hilbert space and fermionizing the projective parameter.}

\vspace*{.5cm}
\begin{flushleft}
~\\
{${}^{\small \mbox\Aries}$\href{mailto:ochandiaq@gmail.com}{ochandiaq@gmail.com}}\\
{${}^{\small \mbox\Pisces}$ \href{mailto:wdlinch3@gmail.com}{wdlinch3@gmail.com}}\\
{${}^{\small \mbox\Libra}$ \href{mailto:vallilo@gmail.com}{vallilo@gmail.com}}
\end{flushleft}

\setcounter{page}0
\thispagestyle{empty}

\newpage
\tableofcontents

\section{Introduction}

The matter spectrum of the ten-dimensional Siegel superparticle \cite{Siegel:1985ys} consists of 10 bosonic space-time coordinates, 16 fermionic space-time superpartners, and 16 canonically conjugate fermionic momentum variables. When extended to the superstring, these variables contribute $10-2\cdot{16}=-22$ to the central charge. This can be canceled by a ten-dimensional pure spinor with 11 holomorphic bosonic degrees of freedom and 11 canonically conjugate momentum variables \cite{Berkovits:2000fe}. As reviewed in section \ref{purespinor}, compactification to $6|8$ dimensions changes the counting to $6-2\cdot 8=-10$, while a six-dimensional pure spinor has only $\frac12\left[2+\frac 62(\frac62 -1)\right]=4$ holomorphic bosonic degrees of freedom and their conjugate momenta \cite{Berkovits:2005hy}. In section \ref{dimred} we will argue that under the dimensional reduction $10|16\to 6|8$, $\mathbf {11}\to \mathbf 6\oplus \mathbf 4\oplus \mathbf 1$ where the $\mathbf 6$ is not covariant under the reduced Lorentz group. Gauge fixing to a ``small'' Hilbert space of variables without this representation is performed in section \ref{reduced} and amounts to passing to six-dimensional projective superspace in which the $\mathbf 1$ plays the role of the projective parameter and the 4 projective constraints are imposed by the six-dimensional pure spinor|a holomorphic Majorana-Weyl spinor of $SU^*(4)$.\footnote{We should comment on the comparison of our proposal to previous work that has appeared in the literature \cite{literature}: Although superficial resemblances with our proposal are guaranteed due to the symmetries of the underlying problems, we emphasize that here we are compactifying the ten-dimensional pure spinor superstring; this requires the understanding of additional degrees of freedom and the recovery of the full set of physical state conditions. Care has been taken to achieve the desired properties by standard methods which do not change the cohomology. These include, the use of unconstrained $U(5)$ variables, demonstration of the decoupling of contractible pairs of worldsheet fields, and fermionization of a curved $\beta\gamma$-system. All of these steps, or something equivalent to them, appear to be necessary for the compactification of the pure spinor string.
}

The field representation theory on these spaces is somewhat singular. It can be regularized by introducing the complex conjugate projective variables and passing to homogeneous coordinates on $\mathbb CP^1$, that is, by lifting to harmonic superspace. This is done in section \ref{harmonic} by introducing the required set of non-minimal variables and extending the algebra of world-sheet currents. We conclude in section \ref{sca} with a proposal for the construction of the $N=2$ worldsheet superconformal algebra.

\section{Pure spinors in non-covariant variables}
\label{purespinor}
We consider the group $Spin(2d)\times U(1)$ generated by
\begin{eqnarray}
N^{mn} = -\frac 12 \lambda^\alpha (\gamma^{mn})_\alpha{}^\beta w_\beta ~~~\mathrm{and}~~~
J= \lambda^\alpha w_\alpha
\end{eqnarray}
where
\begin{eqnarray}
\lambda_\alpha \lambda_\beta = {1\over d!\, 2^d}(\gamma_{m_1\dots m_d})_{\alpha \beta}(\lambda \gamma^{m_1\dots m_d} \lambda) 
\end{eqnarray}
is a $D=2d$-dimensional pure spinor.
The condition is solved in the $SU(d)\times U(1)$ variables by \cite{Berkovits:2005hy}
\begin{eqnarray}
\lambda^\frac d2 = \gamma
~,~
\lambda_{ab}^{\frac d2-2} = \gamma u_{ab}
~,~
\lambda_{abcd}^{\frac d2-4} = \gamma u_{[ab}u_{cd]}
~,~ \dots
\end{eqnarray}
where the $U(1)$-charges are indicated as superscripts. The current for this charge is evidently of the form $N= \frac d2\normal {\gamma \beta} -\normal{u_{ab} v^{ab}}$ where
\begin{eqnarray}
\beta(z_1) \gamma(z_2) \sim \frac1{z_{12}}
~,~
v^{ab}(z_1) u_{cd}(z_2) \sim 2 {\delta^{[a}_c \delta^{b]}_d\over z_{12}^2} .
\end{eqnarray}
The problem is that, in these variables, the original abelian ``ghost-number'' current $J= \gamma \beta$ so that $JN$ has a double pole. Since we will eventually use the $Spin(2d)$ symmetry as a Lorentz symmetry, we wish to avoid this anomaly. One solution is to fermionize
\begin{eqnarray}
\beta = \partial \xi \mathrm e^{-\phi}~~~\mathrm {and} ~~~ \gamma = \eta \mathrm e^\phi
\end{eqnarray}
where
\begin{eqnarray}
\xi(z_1) \eta(z_2) \sim \frac 1{z_{12}} ~~~\mathrm {and}~~~ \phi(z_1) \phi(z_2) \sim -\log z_{12}.
\end{eqnarray}
With these variables and the conditions on ghost number and Lorentz charge of the various variables, we derive that
\begin{eqnarray}
J= - \left(\frac 52 \partial \phi +\frac32 \normal {\eta \xi} \right) ~~~\mathrm {and}~~~
N=\frac d2 \left(\frac32 \partial \phi + \frac52  \normal {\eta \xi } \right)- \normal {u_{ab}v^{ab}}.
\end{eqnarray}

The generator $N=N_a^a$ is the trace of a $U(5)$ generator. It follows that
\begin{eqnarray}
\label{AdjointN}
N_a^b = {\delta_a^b\over 2}\left(\frac 32 \partial \phi + \frac 52 \normal {\eta \xi}\right)- \normal {u_{ac}v^{bc} }.
\end{eqnarray}
The operator product of this current with itself is easily computed to be
\begin{eqnarray}
N_a^b(z_1) N_c^d(z_2) \sim
	{2-d\over z_{12}^2}\delta_a^d\delta_c^b
	+ \frac 1{z_{12}} \left(\delta_a^d N_c^b-\delta_c^b N_a^d\right)(z_2).
\end{eqnarray}
The remaining $d(2d-1) - d^2 = 2 \cdot \frac12d(d-1)$ generators $N^{ab}$ and $N_{ab}$ can be reconstructed from their operator products with $N_a^b$ and among themselves
\begin{eqnarray}
N^{ab}(z_1) N_c^d(z_2) &\sim& {2\over z_{12}} \delta_c^{[a} N^{b]d}(z_2)\cr
N_{ab}(z_1) N_c^d(z_2) &\sim& {2\over z_{12}} \delta_{[a}^d N_{b]c}(z_2)\cr
N_{ab}(z_1) N^{cd}(z_2) &\sim& {2-d\over z_{12}^2} 2\delta_a^{[c}\delta_b^{d]}
	+{2\over z_{12}} \left(\delta_{a}^{[c }N_b^{d]}- \delta_{b}^{[c }N_a^{d]}\right)(z_2).
\end{eqnarray}
This gives
\begin{eqnarray}
\label{HolN}
\vspace{-1cm}
N^{ab} &=& v^{ab}\cr
N_{ab}&=& (d-2)\partial u_{ab} +\normal{u_{ac}u_{bd}v^{cd}}+u_{ab} \left(\frac32\partial \phi+\frac52\normal{\eta \xi}\right).
\end{eqnarray}

\section{Dimensional reduction}\label{dimred}
Starting in $d=5$ and reducing to $d=3$ we break $Spin(10)\to Spin(6)\times Spin(4)$.
For the pure spinors, however, the relevant breaking is $SU(5)\to SU(3)\times SU(2)$.
Then
\begin{eqnarray}
\begin{array}{ccccccc}
u_{ab} & \to & u_{IJ}& &u_{Ia^\prime } && u\\\\
\overline{\mathbf{10}}&=&(\mathbf{3},\mathbf 1)&\oplus&(\bar{\mathbf{3}},\mathbf2)&\oplus&(\mathbf 1,\mathbf{1})
\end{array}
\end{eqnarray}
Together with the singlet $\gamma$, the six-dimensional pure spinor is $\{ \gamma, \gamma u_{IJ}\}$ and the four-dimensional one is $\{\gamma , \gamma u\}$.

One way to see this is to write the Cartan charges of the various spinors $[\pm\pm\pm\pm\pm]$. Without loss of generality, we designate the last two slots as (twice) the values under the Cartan generators of the K3 Lorentz group. Then the $\overline {\mathbf {10}}$ has charge $\frac12$ ({\it i.e.}~three $+$s) and splits as $[***++]\oplus[***\pm\mp]\oplus [***--]$ which are the $(\mathbf 3,\mathbf1)\oplus (\bar{\mathbf 3}, \mathbf 2)\oplus (\mathbf 1, \mathbf 1)$.
The $(\mathbf 3,\mathbf1)$ has the correct ``internal'' structure to combine with $\gamma$ to form the $\mathbf 4$ of $Spin(6)\approx SU(4)$, which is the pure spinor in six dimensions. This spinor satisfies the ``odd-$+$ rule'' in the six-dimensional spin structure and is therefore Weyl. Wick rotating to $Spin(5,1)\approx SU^*(4)$ and switching to $SL(2,\mathbb H)$ notation gives a symplectic-Majorana-Weyl spinor $\lambda^{A{a}}$ where $A=1,\dots,4$ and ${a}=1,2$ is a chiral $SU(2)\hookrightarrow Spin(4)$ index.

The dimensional reduction of the $SU(5)$ generator $(N_a^b)$ in the $SU(2)^\prime$ directions reduces to
\begin{eqnarray}
N_{a^\prime}^{b^\prime} = {\delta_{a^\prime}^{b^\prime}\over 2} \left(\frac32\partial \phi+\frac 52 \normal {\eta \xi} -2\normal {uv} \right) -\normal{u_{K{a^\prime}} v^{K{b^\prime}}}
\end{eqnarray}
This implies that the traceless part
\begin{eqnarray}
\check N_{a^\prime}^{b^\prime} = -\normal{u_{K{a^\prime}}v^{K{b^\prime}}} +\frac12 \delta_{a^\prime}^{b^\prime} \normal{u_{K{c^\prime}}v^{K{a^\prime}}}
\end{eqnarray}
is symmetric upon raising/lowering an index
\begin{eqnarray}
\label{ASDN}
\check N_{{a^\prime}{b^\prime}} = -\normal{u_{K({a^\prime}}v_{{b^\prime})}^{K}}
\end{eqnarray}
Since this generator has only one type of internal $SU(2)$ index, it is of definite duality in $Spin(4)$. If this charge is included in a gauging of a $\mathbb Z_2$ subgroup of $SU(2)^\prime\subset Spin(4)$, the bi-fundamental pure spinor $u_{Ia^\prime}$ will be odd. The resulting description of the pure spinor string on a K3 orbifold was recently worked out in reference \cite{Chandia:2011wd}.

To construct the other $SU(2)$ generator, we recall the relation between 2-forms and their $Spin(4)$ decomposition. The Hodge decomposition of a 2-form $F$ gives $F= F^{(2,0)} + F^{(1,1)}+ F^{(0,2)}$. The Hermitian part is decomposable into a traceless part and a trace. In terms of the K\"ahler form $F^{(1,1)}= \check F^{(1,1)} +  F^{(0,0)} \omega$. The primitive part ($\check F^{(1,1)} \wedge \omega=0$) gives the anti-self-dual $\mathbf 3^\prime$ while the trace combines with the (anti-)holomorphic parts into the self-dual $\mathbf 3$ of $Spin(4)$. In the analysis above, we have found the anti-self-dual generator $(N_{a^\prime b^\prime})$ directly in terms of the bi-fundamental $\mathbf 6$-variables.
Now, since $F^{(0,0)}$ is a trace, we have that, in the original $U(5)$ indices, $N_{1\bar 1}=N_1{}^1=N_2{}^2= N_{2\bar 2}$, a fact which follows by inspection of (\ref{AdjointN}). The other two parts $N_{12}$ and $N_{\bar 1\bar 2}=N^{12}$ are given by (\ref{HolN}).
The self-dual part of the $Spin(4)$ generators is, therefore, given by
\begin{eqnarray}
(N_{ab})&=&
\left( \begin{array}{cc}
N_{1}{}^{1}& N_{12}\\
N^{12} & -N_{1}{}^{1}
\end{array}\right)
\end{eqnarray}
where, on the right-hand-side, we are using the indices of the original $U(5)$ variables; in this case they are (anti-)holomorphic vector indices.
Plugging in the values of the dimensionally reduced generators (\ref{AdjointN}) and (\ref{HolN}) gives\footnote{Here we are taking the $d=2$ part only. The full generator has, in addition, a $d=3$ part and a mixed part coming from the bi-fundamental variables.}
\begin{eqnarray}
\label{SDN}
\hspace{-.5cm}
(N_{ab})=
\left( \begin{array}{cc}
\normal{ u^2v} +u\left(\frac32\partial\phi +\frac52\normal{\eta \xi}\right)
	&\frac12\left( \frac32\partial \phi +\frac52\normal {\eta \xi}-2\normal {uv} \right)\\\\
\frac12\left( \frac32\partial \phi +\frac52\normal {\eta \xi}-2\normal {uv} \right)& v
\end{array}\right).
\end{eqnarray}

At this point we have replaced
\begin{eqnarray}
{Spin(10)\over U(5)} \longrightarrow {Spin(6)\over U(3)}\times {Spin(4)\over U(2)}
\end{eqnarray}
and constructed the pure spinor (which is just an unconstrained symplectic-Majorana-Weyl spinor) of the first factor. The remaining pure spinor degree of freedom $\gamma u$ has charge $\frac12$. We can, of course, consider the two variables separately. Note that since $\gamma u$ has charge $[+++--]$ and $\gamma$ has charge $[+++++]$, it follows that the $-2$ charge of $u$ is entirely in the second factor. Indeed, this second factor
\begin{eqnarray}
{Spin(4)\over U(2)}\approx
	{SU(2)\times SU(2)^\prime\over SU(2)\times U(1)}\approx
	{SU(2)\over U(1)}\approx
	\mathbb CP^1
\end{eqnarray}
is just the parameter space of the harmonic of off-shell superspace with 8 real super-charges. Note also that $u\varepsilon_{\alpha \beta}={\lambda^\frac12_{\alpha \beta}}/{\lambda^\frac52}$ where $\lambda^\frac12_{\alpha \beta}$ is the part of the original $\overline{\mathbf {10}}$ with both legs in the second factor. This combination is ubiquitous in comparisons of the pure spinor with the RNS formalism. The novelty here is that it is uncharged under the six-dimensional Lorentz group, transforms homogeneously under chiral internal rotations ({\it i.e.}~one $SU(2)$ factor in $Spin(4)$), is invariant under the holonomy group of K3 ({\it i.e.}~the other $SU(2)$ factor), and has no ghost charge.

To recapitulate, the dimensionally reduced pure spinors are
\begin{eqnarray}
\begin{array}{cccc}
\lambda^{1} & =  &\gamma & \tiny{\hbox{$[+++,++]$}}\\
\lambda^{1I}&=&\frac12 \epsilon^{IJK}\gamma u_{JK}& \tiny{\hbox{$[+--, ++]$}}\\
\lambda_{Ia^\prime}&=&\gamma u_{Ia^\prime}&\tiny{\hbox{[$++-,+-]$}}\\
\lambda^2&=& \gamma
	u & \tiny{\hbox{$[+++,--]$}}\\
\hat \lambda^I&=&
	\frac12 \epsilon^{IJK}\gamma u u_{JK}
	+\frac12 \epsilon^{IJK}\gamma  u_{Ja^\prime}u_{K}^{a^\prime}&\tiny{\hbox{$[+--,--]$}}\\
\lambda_{a^\prime}&=&\frac12 \epsilon^{IJK}\gamma u_{IJ} u_{Ka^\prime} &\tiny{\hbox{$[---,+-]$}}
\end{array}
\end{eqnarray}
Note that the last two make up the $\mathbf 5$; they are not independent of the $\overline{\mathbf {10}}$.
It will be useful to separate
\begin{eqnarray}
\hat \lambda^I=\lambda^{2I}+\Lambda^I ,
\end{eqnarray}
where
\begin{eqnarray}
	\lambda^{2I} = \frac12 \epsilon^{IJK}\gamma u u_{JK} ~~~\mathrm{and}~~~
	\Lambda^I =\frac12 \epsilon^{IJK}\gamma  u_{Ja^\prime}u_{K}^{a^\prime}.
\end{eqnarray}
Then, the various parts can be reassembled into $Spin(6)\times SU(2)^\prime$ representations as
\begin{eqnarray}
(\lambda^{A1})=\left(\begin{array}{c} \lambda^1\\ \lambda^{1I}\end{array}\right)
~,~
(\lambda^{A2})=\left(\begin{array}{c} \lambda^2 \\ \lambda^{2I}\end{array}\right)
\end{eqnarray}
and
\begin{eqnarray}
(\lambda_{Aa^\prime})=\left(\begin{array}{c} \lambda_{a^\prime}\\ \lambda_{Ia^\prime}\end{array}\right)
~,~
(\Lambda^{A})=\left(\begin{array}{c} 0 \\
	\Lambda^I\end{array}\right).
\end{eqnarray}

The first two spinors can be combined into a $SO(6)$ spinor-valued holomorphic $SO(4)$ spinor as
\begin{eqnarray}
(\lambda^{Aa})=\left(\begin{array}{c} \lambda^{A1}\\ \lambda^{A2}\end{array}\right).
\end{eqnarray}
Note, however, that this latter representation factorizes since $\lambda^{A2}=u \,\lambda^{A1}$: If we define the $SU(2)$ spinor
\begin{eqnarray}
(u^a)= \left(\begin{array}{c} 1\\ u \end{array}\right)
\end{eqnarray}
then we find that the symplectic-Majorana-Weyl spinor can be written as\footnote{We should note that this spinor is not the dimensional reduction of any ten-dimensional spinor since we have had to subtract the non-covariant $\Lambda^A$.}
\begin{eqnarray}
\lambda^{Aa} = u^a \lambda^A.
\end{eqnarray}
This factorization partially solves the dimensionally reduced pure spinor constraints
\begin{eqnarray}
\lambda^{Aa} \lambda_{A}^{a^\prime} = 0~~~ \mathrm {and} ~~~
	\lambda^{Aa}\lambda^{B}_a + \frac12 \epsilon^{ABCD} \lambda_{Ca^\prime}\lambda_D^{a^\prime} =0.
\end{eqnarray}
The spinor $(\lambda_{Aa^\prime})$ is linear in the bi-fundamental ``$\mathbf 6$-variables'' $u_{Ia^\prime}$.
It is easy to see that the remaining pure spinor constraints are satisfied in the form
\begin{eqnarray}
\lambda^A \lambda_{Aa^\prime}=0
~,~
\Lambda^A \lambda_{Aa^\prime}=0
~,~
\lambda_{Aa^\prime}\lambda_B^{a^\prime}=\epsilon_{ABCD}\lambda^C \Lambda^D.
\end{eqnarray}

One now has the option to proceed by writing the pure-spinor-number-1 massless unintegrated vertex operator
\begin{eqnarray}
U= \lambda^{Aa} A_{Aa}+ \lambda_{Aa^\prime} A^{Aa^\prime} +\Lambda^A A_{A2}
\end{eqnarray}
and computing the cohomology of the Berkovits differential
\begin{eqnarray}
\label{Q}
Q= \oint\left( \lambda^{Aa} d_{Aa}  + \lambda_{Aa^\prime}d^{Aa^\prime}+ \Lambda^A d_{A2}\right) .
\end{eqnarray}
This is the type of approach taken in reference \cite{Chandia:2009it} (see also \cite{Chandia:2011wd}). It gives the dimensionally reduced pure spinor constraints on the potentials $A$ in a form which is a central extension of the $N=3$ superspace considered by Sokatchev in his analysis of off-shell, self-dual, $N=4$ Yang-Mills theory \cite{Sokatchev:1995nj}. Checking Bianchi identities up to dimension 2 shows that the conditions to have $N=(1,0)$ target space supersymmetry are that the compactification manifold admit a hermitian metric with a curvature form of definite duality and a holomorphic connection on the gauge bundle with the same duality. In what follows, we will take a different approach which exploits the asymmetry in the compactified pure spinor variables and relates the physical state conditions to the cohomology of pure spinors in lower dimensions.

\section{Reduced Hilbert space}
\label{reduced}
The Berkovits operator (\ref{Q}) has, in order, a term independent of the bi-fundamental $\mathbf 6$-variables, a term linear in them, and a quadratic one. Together with the fact that these variables violate the reduced Lorentz invariance of a K3 compactification, one might be tempted to speculate that it should be possible to describe the compactification consistently without them.

In this section we argue that the cohomology of the Berkovits operator can be computed by use of a simpler differential in a reduced Hilbert space consisting of only those variables not in the bi-fundamental $\mathbf 6$ representation. This follows from a standard argument in homological algebra used in \cite{Berkovits:2001us} in an attempt to relate the analogous operators in the RNS and pure spinor strings.

We proceed by assigning charges $(+1,-1)$ to $(u_{Ia^\prime}, v^{Ia^\prime})$ and $(+2,-2)$ to $(\theta_{Ia^\prime}, p^{Ia^\prime})$. Since the assignment is not required to respect any symmetry {\it a priori}, we will assign charges $(+\alpha, -\alpha)$ to $(\theta_{1a^\prime} , p^{1a^\prime})$.
The Hilbert space
\begin{eqnarray}
C^\bullet= \bigoplus_{n\geq \bar n} C^n
\end{eqnarray}
is graded and bounded below by $\bar n=\bar n(h)$ for any finite conformal weight $h$ because only fields with positive weight carry negative charge.
The world-sheet currents corresponding to the superspace derivatives are
\begin{eqnarray}
d_{Aa} &=& p_{Aa} +i \theta^B_a \partial x_{AB} +i \theta_{Aa^\prime} \partial x^{aa^\prime}
	+\epsilon_{ABCD} \theta^{B}_{a}\theta^{Cb}\partial \theta^D_b\cr\cr
d^{Aa^\prime} &=& p^{Aa^\prime} +i \theta_B^{a^\prime} \partial x^{AB} + i\theta^A_a \partial x^{aa^\prime}
	+\epsilon^{ABCD} \theta_{B}^{a^\prime}\theta_{Cb^\prime}\partial \theta_D^{b^\prime}.
\end{eqnarray}
In what follows, we will ignore the cubic terms in $\theta$. (It can be checked that they do not invalidate the conclusion.)
The Berkovits operator splits into $Q=\sum_n Q_n$ where
\begin{eqnarray}
\hspace{-1.1cm}
\begin{array}{ll}
Q_{-1} = \oint \lambda_{Ia^\prime} p^{Ia^\prime}
	&Q_{1-\alpha} = \oint \lambda_{1a^\prime} p^{1a^\prime}\\
Q_0= \oint \lambda^A d^+_A
	&\\
Q_1=\oint \lambda_{Aa^\prime} \theta^A_a\partial x^{aa^\prime}
	&\\
Q_2= \oint \left[\Lambda^A \left( p_{A2} +\frac i2 \theta^{B1} \partial  x_{AB}\right) + \lambda^I_a\theta_{Ia^\prime} \partial x^{aa^\prime} \right]
	&Q_\alpha = \oint \psi_{aa^\prime} \partial x^{aa^\prime}\\
Q_3= \oint \lambda_{Aa^\prime} \theta^{a^\prime}_I \partial x^{AI}
	&Q_{1+\alpha} = \oint \lambda_{Ia^\prime}\theta^{a^\prime}_1 \partial x^{I1} \\
	&Q_{2+\alpha}= \frac i2 \oint \Lambda^A \theta_A^{a^\prime} \partial x_{2a^\prime}.
\end{array}	
\end{eqnarray}
Here,
\begin{eqnarray}
\label{d+psi}
\hspace{-.5cm}
d^+_{A} = u^a\left(p_{Aa} + i\theta^B_a \partial x_{AB}+\epsilon_{ABCD} \theta^{B}_{a}\theta^{Cb}\partial \theta^D_b\right)
~\mathrm{and}~
\psi_{aa^\prime}=
 \gamma \,u_{a}\, \theta_{1a^\prime}.
\end{eqnarray}
By an argument reviewed presently,
\begin{eqnarray}
H(Q) = \check H^0(Q_0),
\end{eqnarray}
where the second cohomology is computed in a ``small'' Hilbert space in which all the charged variables are zero.

\paragraph{The argument}\hspace{-6pt}goes as follows \cite{Berkovits:2001us}: As already mentioned, the chain complex ends at some $C^{\bar n}$ for any finite-weight vertex operator. Supposing $U= \sum_{n\geq \bar n} U_n$ is in the cohomology of $Q$, $QU=0$ and $\delta U =  \sum_{n\geq \bar n} \alpha_n$. In particular $Q_{-1} U_{\bar n} = 0$ so that $U_{\bar n}$ is independent of $(\theta^{Ia^\prime}, v_{Ia^\prime})$. Then it is easy to see that $U_{\bar n} = u^{Ia^\prime} f_{Ia^\prime} + p_{Ia^\prime} g^{Ia^\prime}$ for some $f\in C^{\bar n-1}$ and $g\in C^{\bar n+2}$ which are independent of $(\theta^{Ia^\prime}, v_{Ia^\prime})$. Therefore $Q_{-1}f=0=Q_{-1}g$ and, consequently,
$U_{\bar n} =\frac1\gamma Q_{-1}(\theta^{Ia^\prime}) f_{Ia^\prime}+Q_{-1}(v_{Ia^\prime}) g^{Iq^\prime}
	=Q_{-1}\left(\frac1\gamma \theta^{Ia^\prime} f_{Ia^\prime}+v_{Ia^\prime} g^{Iq^\prime}\right)$ is exact.\footnote{Note that the factor of $\frac1\gamma$ is necessary since $Q_{-1} \sim \oint \lambda p$ not $\oint up$.
However, since $C^{\bar n-1}=\{0\}$, $f=0$ in this case so that this factor does not appear at this level.
}
Using the gauge freedom $\delta U_{\bar n} = Q_{-1} \alpha_{\bar n +1}$, we can choose $U_{\bar n}$ to be independent of the charged variables or, since a charged vertex independent of the charge variables must vanish, $U_{\bar n}=0$.

This argument works for all $\bar n \leq n< 0$. At $n=0$, we can again choose $U_0$ to be independent of all charged variables but we cannot now use the charge argument to conclude that it vanishes.
Let us denote this choice of gauge by $\check U_0$ so that the condition just stated can be written as $U_0 = \check U_0+Q_{-1} V_1$ for some $V_1$. The descent equation is $0=Q_0 U_0 + Q_{-1} U_1=Q_0 \check U_0 + Q_{-1} (U_1-Q_0 V_1)$. The second term must vanish for suppose it did not. Then it is not hard to see that it depends on the charged variables in chargeless combinations. But since both $Q_0$ and $\check U_0$ are independent of these variables there can be no cancelation between the terms. Therefore, the second term must vanish separately from the first and we arrive at the conclusion that, in this gauge
\begin{eqnarray}
Q_0 \check U_0 = 0 ~~~\mathrm {and}~~~ \delta \check U_0= Q_0 \alpha_0
\end{eqnarray}

At this point the vertex operator takes the form $\check U=\check U_0 + \sum_{n\geq1} U_n$. It is, of course, still annihilated by the complete $Q$ operator. In particular,
\begin{eqnarray}
0&=&Q_{-1} Q\check U
	= \sum_{m\geq0} Q_{-1} \sum_{n=-1}^m Q_{n}\check U_{m-n}
	= \sum_{m\geq0} Q_{-1} \sum_{n=0}^m Q_{n}\check U_{m-n}\cr
&=&Q_{-1} Q_0 \check U_0+ \sum_{m\geq1} Q_{-1} \sum_{n=0}^m Q_{n}\check  U_{m-n} 	
\end{eqnarray}
We use the gauge condition on the first term. Then the remaining sum is over positive values of $m$ where the cohomology of $Q_{-1}$ is trivial. This means that for any $m$, there is an operator $\hat U_{m+1}$ in the equivalence class of $U_{m+1}$ such that $Q_{-1}\hat U_{m+1}=\sum_{n=0}^m Q_{n}\check  U_{m-n}=Q_m\check  U_0+\sum_{n=0}^{m-1} Q_{n}U_{m-n}$. This operator  $\hat U_{m+1}= \frac1{Q_{-1}} \sum^m_{n=0} Q_n \check U_{m-n}$
on the complement of the kernel $\mathrm{ker}_{m+1} Q_{-1}$ and is, therefore, unique up to the usual gauge transformation. This shows that starting with any particular $\check U_0$ in the reduced Hilbert space and the cohomology of $Q_0$, we can construct a representative $U$ of the cohomology of $Q$ uniquely up to gauge transformation. This concludes {the argument}.
\\
~\\

We can restore manifest six-dimensional Lorentz invariance by taking $\alpha=2$. Then according to the result above, we are computing the cohomology of $Q_0=q$, where
\begin{eqnarray}
q=\oint \lambda^A d^+_A,
\end{eqnarray}
imposes the projective constraint on six-dimensional superfields. Alternatively, we can take $\alpha=0$.\footnote{The conclusion holds for any $\alpha$ but at certain values one of the $\alpha$-dependent $Q_*$s gets a vanishing charge. The candidates are $\alpha = 1$, $0$, $-1$, and $-2$. These are relevant to the description of {\em four}-dimensional compactifications that are, respectively, chiral, self-dual, anti-holomorphic in the internal space, and anti-holomorphic in space-time. 
}
In this case $Q_0= q+ \delta$ where
\begin{eqnarray}
\delta= \oint  \psi_{aa^\prime} \partial x^{aa^\prime}.
\end{eqnarray}
Since $0=q^2=\{q,\delta\}=\delta^2$, the the charge-0 cohomology becomes the relative cohomology
\begin{eqnarray}
\check H^0(Q_0) = H(q; H(\delta))= H(\delta;H(q))
\end{eqnarray}
of projective superfields $H(q)$ with values in the cohomology $H(\delta)$. For ghost-number-1 vertex operators
\begin{eqnarray}
U= \psi^{aa^\prime} A_{aa^\prime},
\end{eqnarray}
this describes self-dual gauge fields on K3 \cite{Berkovits:2004ib} with projective superfield wave functions.

\section{Non-minimal variables and harmonic constraints}\label{harmonic}
In the harmonic superspace formalism \cite{Galperin:2001uw}, the vector multiplet emerges as a connection, not of the potential $A^+_A$, which is pure gauge, but as a connection of an auxiliary harmonic operator $D^{++}= \epsilon_{ab} u^{+a} \partial/\partial u^{-b}$. The role of this operator is to control the Laurent expansion of a harmonic superfield of definite harmonic charge by imposing constraints on its $u^-$-dependence. For example, a field $\Phi^+(u^\pm)$ with harmonic charge $+1$ and satisfying $D^{++} \Phi^+=0$ is simply a hypermultiplet $\Phi^+ = u^{+a} \varphi_a$.

In projective superspace, the variable $u^-$ is related to the complex conjugate $\bar u$ of the projective variable $u$. Since in this description all fields are holomorphic, the condition $D^{++}=0$ is automatic. There are complications in the projective formalism due to this strict holomorphicity which can be alleviated by introducing the conjugate variable $\bar u$ as a regulator.

In this section we will treat the condition $D^{++}=0$ of projective superspace as a gauge fixing condition on harmonic superspace. This is equivalent to introducing the $u^-_a=\bar u_a$ variable into the formalism as a non-minimal field and introducing the BRST operator $s$ and ghost $\kappa^-_a$ such that $s(u^-_a)=\kappa^-_a$.

In anticipation of a complication we would encounter later, we take the opportunity here to also introduce a non-minimal field corresponding to a homogeneous coordinate on $\mathbb CP^1$: We may think of the holomorphic parameter $u$ as the ratio ${u^{+2}\over u^{+1}}$ of homogeneous coordinates $[\,u^{+1}\!:\! u^{+2}\,]$ and replace our old $u^a \to u^{+a} = u^{+1}u^a$. Every instance of a ${}^+$ heretofore should be un-gauge-fixed in this way. For example $d^+_A =d_{A1}+ud_{A2} \to u^{+1}(d_{A1}+ud_{A2})=u^{+a}d_{Aa}$. Of course, this procedure requires a modification of the BRST operator so that, additionally, $s(u^{+1})=\kappa^+$.

To implement these changes covariantly, we introduce the momenta conjugate to the fields and ghosts
\begin{eqnarray}
v^-_a (z_1) u^{+b}(z_2)\sim \frac{\delta_a^b}{z_{12}}
~~~,~~~
	\delta^- (z_1) \kappa^+(z_2)\sim \frac{1}{z_{12}}
\end{eqnarray}
\begin{eqnarray}
v^{+a} (z_1) u^-_b(z_2)\sim \frac{\delta_b^a}{z_{12}}
~~~,~~~
	\delta^{+a} (z_1) \kappa^-_b(z_2)\sim \frac{\delta_b^a}{z_{12}}
\end{eqnarray}
and modify the Berkovits operator by replacing
\begin{eqnarray}
\label{Qlift}
Q_0\to Q_{\hat 0}=Q_0+ \oint u^{+1}\delta^-
+ \oint \kappa^-_a  v^{+a} .
\end{eqnarray}
This extension of the Hilbert space introduces new $SU(2)_{-4}$ currents\footnote{As indicated, the level of this algebra is $-4$. This implies that the 0-mode normalization requires a compensation of the background charge of $+4$. This is the charge carried by the analytic subspace, indicating that the 0-mode normalization corresponds to the analytic measure. (See also reference \cite{Chandia:2011wd}.)}
\begin{eqnarray}
\label{su2currents}
d^{++} =\sqrt2 \epsilon_{ab} u^{+a}v^{+b}
,~
d^0= \normal{u^{+a}v^-_a} - \normal{u^-_a  v^{+a}}
,~
d^{--} = \sqrt2\epsilon^{ab} u^-_a v^-_b,
\end{eqnarray}
and a second harmonic derivative
\begin{eqnarray}
d^-_A= \epsilon^{ab} u^-_b d_{Aa}
\end{eqnarray}
needed to close the algebra. The string fields in the minimal description are now lifted to the space including the new variables but are constrained to lie in the cohomology of the new differential $Q_{\hat 0}$.

The usual ambiguities in the extension of the differential (\ref{Qlift}) have been fixed by requiring that the non-minimal terms anti-commute with $Q_0$ and that the new differential $Q_{\hat 0}$ commute with $d^{++}$. The remaining currents do not commute with the differential. (Instead, $S_{\hat 0}=[d^{--},Q_{\hat 0}]$ defines a new differential which anti-commutes with $Q_{\hat 0}$ and forms a spin-$\frac12$ representation of the algebra (\ref{su2currents}) with it.)

We will say that a vertex operator $\Phi$ has harmonic charge $q$ if\footnote{It is this definition which would have been complicated by not introducing the homogenous coordinates for the projective sphere.}
\begin{eqnarray}
d^0(z_1) \Phi(z_2) \sim \frac1{z_{12}} D^0 \Phi(z_2) = \frac q{z_{12}}\Phi(z_2).
\end{eqnarray}
Recovery of the minimal prescription obtains by projecting back down using the condition that the fields have no poles with $d^{++}$:
\begin{eqnarray}
0= d^{++} (z_1) \Phi(z_2) \sim \frac1{z_{12}} D^{++}\Phi(z_2).
\end{eqnarray}
For a vertex operator with definite harmonic charge, this condition truncates the expansion, thereby putting the vertex operator on-shell.

Next, we want to deform the theory with a background gauge field. The minimal prescription $U=\lambda^A A^+_A$ leads to a harmonic, charge-1 potential which, as already mentioned, is pure gauge: $A^+_A=-\oint d^+_A \Omega=-D^+_A \Omega $ for a bosonic, harmonic-charge-0 ``prepotential''  string field $\Omega$. Removing this deformation is equivalent to performing the transformation
\begin{eqnarray}
\mathcal O \mapsto \mathrm e^{-\Omega} \mathcal O \mathrm e^{\Omega}
\end{eqnarray}
on the entire conformal field theory. This, however, introduces the connection
\begin{eqnarray}
D^{++} \mapsto D^{++} + V^{++}
\end{eqnarray}
with $V^{++}=\oint d^{++} \Omega= D^{++}\Omega$. While it is possible to remove the connection from $D^{++}$ or $D^+_A$, it is not possible to remove it from both simultaneously.

The prepotential $\Omega$ is a real harmonic function of charge 0. Due to the absence of poles in the $d^{++}$ operator product with $d^+_A$ and itself, the prepotential has a gauge transformation of its own: Under $\delta \Omega = \Lambda+\bar \Lambda$ with parameters constrained by $D^{++}\Lambda=0=D^{++} \bar \Lambda$, both $V^{++}$ and $A^+_A$ are invariant. It is easy to see that such harmonic superfields are polar so that we recover the projective superfield description of the gauge multiplet.

\paragraph{The integrated vertex operator} The integrand of Siegel's integrated vertex operator \cite{Siegel:1985xj} can be rewritten in an interesting way using the harmonic integral:
\begin{eqnarray}
I_\mathrm{Siegel} &=& \partial \theta^{Aa} A_{Aa} +\frac12 \Pi^{AB}A_{AB} + d_{Aa} W^{Aa}\cr
	&=&\oint \mathrm d^2u \left(
		\partial \theta^{+A} A^-_{A} +\frac12 \Pi^{AB}A_{AB} + d^-_{A} W^{+A}
	\right).
\end{eqnarray}
The harmonic superspace constraints are solved by
\begin{eqnarray}
A^-_A = D^+_A V^{--}
,~
A_{AB}=(D^+)^2_{AB} V^{--}
,~
W^{+A} = (D^+)^{3A} V^{--}
\end{eqnarray}
which implies that
\begin{eqnarray}
\hspace{-.7cm}
I_\mathrm{Siegel}=\oint \mathrm d^2u \left(
		D^{++}+ \partial \theta^{+A} D^+_{A} +\frac12 \Pi^{AB}(D^+)^2_{AB} + d^-_{A}(D^+)^{3A}
	\right)V^{--}.
\end{eqnarray}
Indeed, we can go one step further by introducing the natural operator
\begin{eqnarray}
\nabla^{++} =D^{++}+ \partial \theta^{+A} D^+_{A} + \Pi^{AB}(D^+)^2_{AB} + d^-_{A}(D^+)^{3A} + d^{--} (D^+)^4,
\end{eqnarray}
and proposing to change the integrated vertex operator to\footnote{Siegel recently suggested the extension of the set of worldsheet fields by ``dual isotropy coordinates'' and their conjugate momenta \cite{Siegel:2011sy}. In such an extension, precisely this type of modification of the vertex operator appears (although here the operator is composite).}
\begin{eqnarray}
\label{IntVertex}
I&=&\oint \mathrm d^2u \, \nabla^{++}V^{--}.
\end{eqnarray}
This differs from $I_\mathrm {Siegel}$ by a term $\oint d^{--}F^{++}$ where
\begin{eqnarray}
F^{++} = (D^+)^4V^{--}
\end{eqnarray}
is the superfield extension of the auxiliary field $F_{ab}|_{\theta=0}$:
\begin{eqnarray}
D_{Aa} W^{Bb} =  \delta_a^b (\sigma^{mn})_A{}^BF_{mn}+ \delta_A^BF_a{}^b.
\end{eqnarray}

The vertex operator (\ref{IntVertex}) is the string theoretic analogue of an operator introduced by Siegel in a slightly different context to solve the analyticity condition $[ \mathcal D^+_A, \mathcal D^{++}] =0$ in terms of the four-dimensional, $N=2$ supergravity prepotential $U=U(x, \theta^\pm)$ \cite{Siegel:1995px} (see also \cite{Siegel:2011sy}). Indeed, following Siegel's result, we can show that the integrand can be written as
\begin{eqnarray}
\nabla^{++} V^{--} =
	D^{++}V^{--}
	+ \frac1{4!} \epsilon^{ABCD} d^+_A\left( d^+_B\left( d^+_C\left(d^+_D\left( V^{--}d^{--}\right)\right)\right)\right).
\end{eqnarray}
This gives the integrated vertex operator for the gauge field in the same form as that which comes from the six-dimensional hybrid formalism \cite{ProjHybrid}
\begin{eqnarray}
I = \frac1{4!} \oint \mathrm d^2 u\, \epsilon^{ABCD} d^+_A\left( d^+_B\left( d^+_C\left(d^+_D\left( V^{--}d^{--}\right)\right)\right)\right).
\end{eqnarray}
We will use this observation in the next section to construct a twisted $N=2$ superconformal algebra which gives the integrated vertex operator in this form.

\section{Relation to the hybrid formalism}
\label{sca}
Although the harmonic and projective superfield formalisms are equivalent \cite{Kuzenko:1998xm}, the manner in which the physical state conditions are implemented in these cases is quite different. (Since superstrings give on-shell representations, this is crucial for us.) In both cases, it is sufficient to demonstrate that the vertex operator is holomorphic/$u^-$-independent and has terminating expansion. In the harmonic case this is implemented by the condition that the vertex have well-defined harmonic charge and that it be annihilated by (some power of) $D^{++}$. In the projective case, the vertex operator is assumed to be homomorphic from the outset and the truncation is equivalent to the condition that the vertex operator be entire on $\mathbb CP^1$ \cite{Kuzenko:1998xm}.

In the projective hybrid formalisms \cite{ProjHybrid}, it is not assumed from the outset that the fields are entire. It is necessary, therefore, to be able to define conditions which implement the truncation. One such condition is the world-sheet chirality of the compactification-dependent vertex operators which bounds the expansion in the projective parameter below. Contact can be made with the hybrid formalism by fermionizing the projective parameter as
\begin{eqnarray}
u=-\mathrm e^{-\rho-i \sigma}
\end{eqnarray}
where the operator products of the chiral bosons are defined to be
\begin{eqnarray}
\rho(z_1) \rho(z_2) \sim - \log z_{12} ~~~\mathrm{and}~~~ \sigma(z_1) \sigma(z_2) \sim - \log z_{12}.
\end{eqnarray}
This splitting implies a gauge symmetry $(\rho, \sigma)\mapsto(\rho+\pi, \sigma+i\pi)$ which is fixed by introducing the constraint
\begin{eqnarray}
J=\partial(\rho +i\sigma).
\end{eqnarray}

Now it is easy to define a chirality condition which implies that the vertex has no negative powers of $u$: It suffices to define
\begin{eqnarray}
G^-=\mathrm e^{-i\sigma}
\end{eqnarray}
and to impose that the vertex operator have no simple pole with it. The implied superconformal algebra is completed by
\begin{eqnarray}
G^+=-\frac1{4!} \epsilon^{ABCD} d_A^+(d_B^+(d_C^+(d_D^+(\mathrm e^{2\rho+3i\sigma})))) ,
\end{eqnarray}
and its operator product with $G^-$. These are the generators of the twisted $N=2$ super-conformal algebra of the first reference in \cite{ProjHybrid}.\footnote{Unfortunately, our conventions differ by the interchange of $\theta^{A1}\leftrightarrow \theta^{A2}$.}
The resulting stress-energy tensor is missing the contribution of the six-dimensional pure spinor \cite{Berkovits:2000fe}. It can be included by shifting
\begin{eqnarray}
G^-\to G^- +w_A\partial \theta^{A1},~
G^+\to G^+ + \lambda^A d^+_A,~
J\to J+\normal{\lambda^A w_A}.
\end{eqnarray}
That the extension of the $G^-$ current has trivial operator product with the first term in $G^+$ is due to the fact that the expansion of the latter reveals that it has no $d_{A1}$-dependence.

It remains only to understand the internal part of the vertex operator. For the compactification-independent vertex operators this is trivial: The unintegrated vertex operators are proportional to the identity operator with six-dimensional projective superfield wave functions $U$. Their integrated version is given by $I=\int G^+(G^-(U))$. The compactification-dependent vertex operators are worldsheet (anti-)chiral analogues of these. Since the algebra is twisted, the vertex operator is $A^{aa^\prime} \psi_{aa^\prime}$, where the internal operator is given in (\ref{d+psi}).

\section{Conclusion}
We have compactified the pure spinor formalism on a K3 manifold keeping manifest, as much as possible, the symmetries of the theory. This process requires a cohomological argument allowing the reduction to a small Hilbert space of worldsheet variables. The resulting complex is trivial when the string fields are completely unconstrained unless some regularization is used. This problem is familiar from projective superspace.\footnote{A string-theoretic analogue of this problem was recently emphasized by Berkovits in the case of the twistor-string description of self-dual Yang-Mills in four dimensions \cite{Berkovits:2004ib} where the analogue of the $u$ variable enters as a holomorphic coordinate on $\mathbb C$. The proposed resolution is to restrict the gauge transformation to be well-defined at $u=\infty$, that is to say, $u$ is a holomorphic coordinate on $\mathbb CP^1$ and gauge parameter superfields must be entire as functions on the sphere. This argument was then applied to the ten-dimensional string using the analogous parameters $u_{ab}$ as projective pure spinors in an attempt to relate the pure spinor fornalism to an $N=4$ topological string \cite{Berkovits:SimonsTalk}.
}
We propose to use harmonic superspace as a regulator and introduce the necessary worldsheet variables as non-minimal fields. This is precisely analogous to the non-minimal extension of the ten-dimensional pure spinor \cite{Berkovits:2005bt}. Indeed, the more logical course of action would have been to introduce these variables {\it ab initio} and remove the extraneous ones using a suitable modification of the homological argument. This, we expect, would lead to the same result at the cost of considerable complication of the argument of section \ref{reduced} and the presentation in general.

Although the harmonic parameters were introduced solely to simplify the description of the cohomology, their presence provides hints to the connection with the RNS string. In particular, they allow us to rewrite the integrated vertex operator in a form suitable for comparison to the hybrid formalism and their algebra has an anomaly leading to a background charge in the 0-mode normalization rule. Consequently, we propose that the (space-time part of the) $N=2$ worldsheet superconformal algebra can be constructed simply by returning to the small Hilbert space and fermionizing the projective parameter. In this description the correct 0-mode measure is the projective superspace measure (which corresponds to that on the analytic subspace of harmonic superspace). It consists of an integral over 4 of the 8 $\theta$s and an additional integral over $u$, in agreement with the charge anomaly in the $SU(2)_{-4}$ algebra generated by the currents (\ref{su2currents}) and the existence of the additional holomorphic variable \cite{Chandia:2011wd}.

\section{Acknowledgements}
We thank Gabriele Tartaglino-Mazzucchelli for discussions about superspaces. W{\sc dl}3 is supported by {\sc fondecyt} grant number 11100425 and {\sc dgid-unab} internal grant  DI-23-11/R. The work of BCV is partially supported by {\sc dgid-unab} internal grant DI-22-11/R.


\end{document}